\documentclass[pre,twocolumn,superscriptaddress,floatfix,nofootinbib]{revtex4-1}  

\usepackage[paperwidth=210mm,paperheight=297mm,centering,hmargin=2cm,vmargin=2.5cm]{geometry}

\usepackage[pdftex]{graphicx}
\usepackage{amsmath,bm}
\usepackage{natbib}
\usepackage{MnSymbol}
\usepackage{csquotes}
\usepackage{mathtools}

\usepackage{hyperref}
\usepackage{xcolor} 
\usepackage{array}
\usepackage{pgfplots}
\usepackage{graphicx}
\usepackage[left,modulo,pagewise]{lineno}
\usepackage{dsfont}

\definecolor{bluemoi}{rgb}{0.25,0.50 ,0.75} 

\hypersetup{
  bookmarksopen=true,
  pdftitle=" ",
  pdfauthor=" ", 
  pdfsubject=" ", 
  pdftoolbar=true, 
  pdfmenubar=true, 
  pdfhighlight=/O, 
  colorlinks=true, 
  pdfpagemode=UseNone, 
  pdfpagelayout=SinglePage, 
  pdffitwindow=true, 
  linkcolor=bluemoi, 
  citecolor=bluemoi, 
  urlcolor=bluemoi 
}

\makeatletter
\renewcommand\@biblabel[1]{#1} 

\renewcommand\newblock{\hskip .11em\@plus.33em\@minus.07em}
\makeatother

\setcitestyle{authoryear,round}
\bibpunct{(}{)}{;}{a}{,}{}

\makeatletter
\newcommand{\removeperiod}{\@ifnextchar.{\@gobble}\relax}
\makeatother 

\makeatletter
\renewcommand{\figurename}{\sf \textbf{Figure}}
\renewcommand{\thefigure}{\arabic{figure}}
\renewcommand{\fnum@figure}{\sf\textbf{\figurename}~\textbf{\thefigure}}
\renewcommand{\tablename}{\sf\textbf{Table}}
\renewcommand{\thetable}{\arabic{table}}
\renewcommand{\fnum@table}{\sf\textbf{\tablename}~\textbf{\thetable}}
\makeatother

\begin{document}

\title{Coupling in situ and remote sensing data to assess $\alpha$- and $\beta$-diversity over biogeographic gradients} 

\author{Maxime Lenormand}
\thanks{Corresponding authors: maxime.lenormand@inrae.fr}
\affiliation{TETIS, University of Montpellier, AgroParisTech, Cirad, CNRS, INRAE, Montpellier, France}

\author{Jean-Baptiste F{\'e}ret}
\affiliation{TETIS, University of Montpellier, AgroParisTech, Cirad, CNRS, INRAE, Montpellier, France}

\author{Guillaume Papuga}
\affiliation{AMAP, University of Montpellier, CIRAD, CNRS, INRAE, IRD, Montpellier, France}

\author{Samuel Alleaume}
\affiliation{TETIS, University of Montpellier, AgroParisTech, Cirad, CNRS, INRAE, Montpellier, France}

\author{Sandra Luque}
\affiliation{TETIS, University of Montpellier, AgroParisTech, Cirad, CNRS, INRAE, Montpellier, France}

\begin{abstract} 
The mapping of plant biodiversity represents a fundamental stage in establishing conservation priorities, particularly in identifying groups of species that share ecological requirements or evolutionary histories. This is often achieved by assessing different spatial diversity patterns in plant population distributions. In this paper, we present two primary data sources crucial for biodiversity monitoring: in situ measurements from botanical observations and remote sensing (RS). In situ methods involve directly collecting data from specific sites, providing detailed insights into ecological patterns but often constrained by resource limitations. Integrating in situ and RS data highlights their complementary strengths, which depend on factors such as study scale, resolution, and logistical feasibility. While in situ approaches are characterized by precision, RS offers efficiency and extensive, repeated coverage. This research integrates in situ and RS data to analyze plant and spectral diversity across France at a spatial resolution of 5 km, encompassing over 23,000 grid cells. We employ four established diversity metrics leveraging the spatial distribution of 6,650 plant species and 250 spectral clusters (derived from MODIS data at a 500-meter resolution). Through bioregionalization network analysis combining these data sources, we identified five distinct bioregions that capture the biogeographical structure of plant biodiversity in France. Additionally, we explore the relationship between plant species diversity and spectral cluster diversity within and between these bioregions, offering novel insights into the spatial dynamics of plant biodiversity.
\end{abstract}

\maketitle

\section*{Introduction}

Ecosystems are dynamic complexes of plant, animal, and micro-organism communities and their abiotic environments that interact as a functional unit. Given the unprecedented nature of climate change, biological diversity is facing serious threats on a global scale \citep{Cardinale2012}. Indeed, at least 25\% of species are threatened with extinction, and most indicators of the state of nature suggest that they are also declining rapidly \citep{Diaz2019,Tickner2020}. These declines can have dramatic and far-reaching negative impacts on human well-being and environmental security \citep{Newbold2019}. Therefore, ensuring accurate biodiversity mapping and monitoring at appropriate spatio-temporal resolutions is crucial for identifying areas of high conservation priority \citep{Jetz2019,Cavender2022,Skidmore2021}. In situ and remote sensing (RS) information show strong complementarity for such tasks.

The relative importance of in situ and remote sensing (RS) methods may be influenced by multiple factors, including the scale of the study \citep{Rocchini2011} and the logistical cost of data acquisition for each type. Methods relying on in situ data collection are well-suited to fine-scale analyses and generally provide more accurate ecological information, such as species identification \citep{Vecera2019, Sabatini2022, Fernandez2013, Zellweger2017}, but they can be resource-intensive and time-consuming. Conversely, RS is advantageous for spatial exhaustivity, upscaling local observations, and frequent revisits \citep{Rocchini2014, Ustin2021, Luque2018, Pettorelli2018, Tickner2020, Rocchini2021, Chraibi2021}. RS enables rapid and broad-scale observation of the Earth's surface but may lack the precision of ground-based observations for localized analyses.

The production of remotely sensed diversity metrics can be achieved through various methods \citep{Luque2018, Ustin2021, Wang2019}. The choice of a method depends on the type of ecosystem, the available RS data, and the dimension of diversity of interest. Some remotely sensed diversity metrics theoretically allow for the assessment of ecologically relevant metrics with little to no supervision. For example, methods based on the Spectral Variation Hypothesis (SVH) assume that remotely sensed spatial heterogeneity reflects the heterogeneity of Earth's surface in terms of ecological entities \citep{Palmer2002,Asner2017,Torresani2024}. These entities may correspond to species, habitats, or biomes, depending on the ecosystem under study and the RS data type. High spatial resolution imagery, with submetric resolution, enables the differentiation and delineation of fine-scale ecological entities, such as individual trees in a forest ecosystem. In contrast, moderate spatial resolution imagery, ranging from 100 m to 1 km, provides an integrated view of Earth's surface composition and processes, such as the dominant community, ecosystem, or biome within the pixel, or related phenology. Here, we will focus on this latter type of information, analyzing the vegetation seasonality derived from MODIS data at 500-meter spatial resolution. For all RS diversity metrics requiring little to no supervision, validation with in situ measurements that are compatible with RS analysis frameworks remains crucial. 

The relationship between diversity metrics measured with in situ and RS data has been investigated \citep{Rocchini2011, Rocchini2014, Marzialetti2021} for various components of biodiversity, including taxonomic, functional, and phylogenetic diversity. When focusing on taxonomic diversity, $\alpha$-diversity (diversity within a community) and $\beta$-diversity (changes in species composition between communities, partitioned into turnover and nestedness components) are two key elements of biodiversity assessments \citep{Whittaker1960, Whittaker1972, Magurran2021}. The methods used to gather data, whether through in situ or RS approaches, can significantly influence the resulting assessments.

Field data have been used to measure $\alpha$-diversity and $\beta$-diversity through various methods across different ecosystems and geographic regions \citep{Vecera2019,Sabatini2022,Fernandez2013,Zellweger2017}. However, most empirical studies examining the relationship between $\alpha$-diversity and $\beta$-diversity assessed using in situ and RS data rely on limited datasets \citep{Chraibi2021,Fassnacht2022}, either taxonomically (restricted to a specific group of species) or spatially (local-scale analyses). A recent study conducted in the Czech Republic demonstrated a significant relationship between species richness and functional diversity from in situ observations and spectral diversity derived from RS data for biodiversity assessment \citep{Perrone2023}. The same study emphasized that the strength of this relationship is context-dependent \citep{Perrone2023}. \citet{Fassnacht2022} identified four contextual factors influencing the relationship between plant biodiversity and spectral diversity: (1) the types of vegetation considered, (2) the scale of the study and sensor resolution, (3) the chosen metric, and (4) changes in reflectance over time due to phenology or other temporal effects. Seasonal variations pose challenges, as they can significantly drive spectral variation at a given location, even during the main growing season. Nevertheless, incorporating seasonal variation through an RS time series provides a valuable source of information on the vegetation's functional and floristic diversity. The relationship between floristic diversity and moderate-resolution RS data, such as MODIS imagery, is not direct: surface reflectance integrated over 1 km$^2$ does not inherently convey information about species richness. However, deeper analyses integrating temporal (phenology) and spatial (heterogeneity or dissimilarity with local and distant neighborhoods) information can yield ecological insights. For instance, the relationship between floristic similarity and land surface phenology similarity has been utilized to assess diversity patterns using MODIS data, either alone or fused with Landsat data \citep{Vina2012,Vina2016}.

In this study, we used in situ and RS data to measure the diversity of plant species and spectral clusters integrated over a 5-km grid covering the metropolitan French territory ($> 23,000$ grid cells). Specifically, we relied on four diversity metrics (species richness, Shannon diversity, Simpson dissimilarity and the turnover component of the Bray-Curtis dissimilarity) to assess $\alpha$-diversity and $\beta$-diversity among grid cells. 

Plant species were extracted from the French National Botanical Conservatory database. The seasonality of vegetation was defined based on a Normalized Difference Vegetation Index (NDVI) time series acquired from the MODIS satellite at 500-meter resolution and discretized using a clustering approach to produce what we define as \emph{spectral clusters}. As mentioned above, the vegetation phenology captured by spectral clusters within 500-meter pixels provides an integrated view of the habitats and land covers included in the pixel. With approximately 50\% of the metropolitan French territory covered by forests, semi-natural areas, wetlands, pastures, and grasslands, we assumed that vegetation phenology integrated over such spatial resolution could provide information related to plant diversity reported from fine-scale inventory databases. 

We first assessed the consistency of diversity metrics derived from plant species and spectral clusters at a national scale across the metropolitan French territory. Subsequently, we conducted a detailed analysis of these metrics at a regional scale, emphasizing the role of spatial context in comparing the two data sources. To achieve this, we adopted a bioregionalization approach \citep{Wallace1876}, dividing the French territory into regions with similar species composition. The underlying principle is to identify regions with minimal internal heterogeneity in taxonomic composition while ensuring maximum differentiation between them \citep{Kreft2010}. More specifically, we utilized a network-based clustering approach \citep{Antonelli2017, Lenormand2019} to delineate bioregions by identifying a minimal common spatial structure shared by plant species and spectral clusters, thereby determining the most suitable regional scale for analysis. Finally, we investigated how plant species diversity and spectral cluster diversity relate to one another both within and across the defined bioregions.

\section*{Materials and Methods}

\subsection*{Datasets and Study Area}

\begin{figure}[!ht]
	\centering 
	\includegraphics[width=\linewidth]{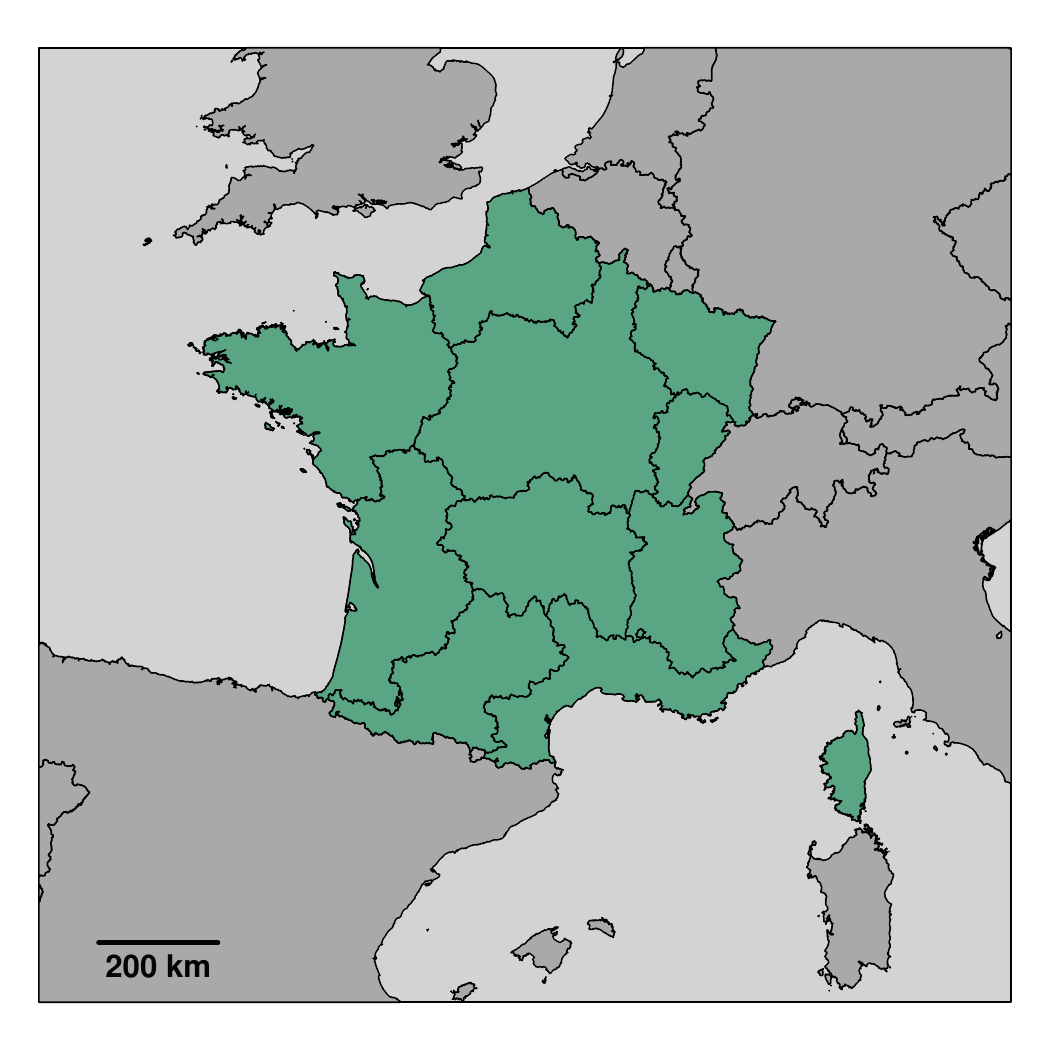}
	\caption{\textbf{Map of the France metropolitan area.} The map shows the National Botanical Conservatories' boundaries.}
	\label{Fig1}
\end{figure}

One of the main goals of the analysis proposed in this paper is to assess the level of similarity derived from two different sources of information to identify biogeographic gradients at a regional/national scale. This analysis was carried out on the metropolitan French territory, including Corsica, which covers an area of 552,000 km$^2$ (Figure \ref{Fig1}). The map of France was divided into $N = 23,060$ square cells of 5-km spatial resolution to define a spatial scale of analysis common for both in situ and RS data.

Information on plant species was extracted from the SIFlore database\footnote{\url{https://siflore.fcbn.fr}}, which was compiled by the Federation of the National Botanical Conservatories in 2016 \citep{Just2015}. It integrates in situ data from the French National Botanical Conservatories, the national reference institute for flora monitoring. It contains information on the presence and abundance of $P=6,650$ plant species. Figure S1 in the Appendix provides the number of observations and plant species per grid cell.

RS data used for this study were based on the Terra Moderate Resolution Imaging Spectroradiometer (MODIS) Vegetation Indices (MOD13A1 Version 6)\footnote{\url{https://modis.gsfc.nasa.gov/data/dataprod/mod13.php}} at 500 meter spatial resolution. The data were obtained from the United States Geological Survey (USGS) data catalog and downloaded using the R package \textit{MODIStsp} \citep{Busetto2016}. This study focused on the NDVI time series acquired in 2010. Once downloaded, each image product was checked visually and based on pixel product quality. Images showing artifacts or poor quality were discarded. The analysis used 15 NDVI products acquired between February 18th 2010 and October 16th 2010. The method implemented in the R package \textit{biodivMapR} \citep{Feret2020} was applied to the NDVI products to produce spectral species maps. This method was initially developed using high spatial resolution airborne imaging spectroscopy to assess taxonomic diversity over tropical rainforest ecosystems. The main principle of the method, applicable to any raster data, is to apply a k-means clustering analysis over an image after appropriate preprocessing and to consider each cluster as an ecological entity defined as a \emph{spectral cluster}. The principle of the method is detailed in \citet{Feret2014} and \citet{Feret2020}, and we will only describe the main steps here. First, we performed dimensionality reduction of the NDVI time series after applying a principal component analysis. The component selection was performed based on visual interpretation of the resulting components to exclude residual artifacts and noise. A k-means clustering was then applied to the selected set of components. These clusters, defined as spectral species in the original publication from \citet{Feret2014}, do not allow species discrimination given the moderate spatial resolution of imagery data. This is why we decided to use the term spectral cluster. In this study, we defined 250 spectral clusters to account for the diversity of seasonal patterns observed on land surface over the French territory, with a fraction of these patterns probably mainly driven by anthropic activities. At the end of the process, we obtained information on the distribution of spectral clusters in each 25 km$^2$ grid cell, including \enquote{presence} and \enquote{abundance} (see Figure S1 in Appendix for more details).

\subsection*{Measuring $\alpha$-diversity and $\beta$-diversity}

\subsubsection*{$\alpha$-diversity metrics}

This study focuses on two well-known $\alpha$-diversity metrics \citep{Hill1973}. First, we consider a 0-order index through the observed species richness per cell, which is computed as the number of different plant species and spectral clusters in each cell (noted as $R^P$ and $R^S$, respectively). Second, we computed a first-order diversity index using the normalized Shannon diversity index \citep{Shannon1948}, which is computed as follows for a given cell:
\begin{equation}
	H^P = -\frac{1}{ln(P)}\sum_{p=1}^P F_{p}ln(F_{p})
	\label{HP}
\end{equation}
Where $F_{p}$ is the fraction of observations associated with the plant species $p$ among all the plant species observations in the considered cell. The normalizing factor, represented by the expression $ln(P)$, corresponds to the uniform case, in which each plant species has the same number of observations. With this normalization, the Shannon diversity index ranges from 0, which represents a situation in which only one plant species has been observed, to 1, which represents a situation in which all species have the same number of observations.

The same formula can be used to compute the Shannon diversity index $H^S$ for the spectral clusters by replacing $s$ with $p$ and $S$ with $P$ in Equation \ref{HP}.

\subsubsection*{$\beta$-diversity metrics}

$\beta$-diversity, which represents changes in species composition among assemblages, can arise from two distinct phenomena: spatial turnover of species and nestedness of assemblages \citep{Baselga2010}. These phenomena result from contrasting processes: species replacement and species loss, respectively. Here, we chose to focus on the turnover component of two $\beta$-diversity indices, with the objective of describing spatial turnover without the influence of richness gradients.

The first index is the Simpson dissimilarity index \citep{Simpson1943}, the turnover component of the S{\o}rensen index \citep{Baselga2010}, calculated between a given pair of distinct cells with the following formula:  
\begin{equation}
	\beta^P_{SIM} = \frac{min(b_P,c_P)}{a_P+min(b_P,c_P)}
	\label{betasim}
\end{equation}
Where $a_P$ is the number of plant species in common between the two considered cells, $b_P$ the number of plant species that are present in the first cell but not in the second and $c_P$ the number of plant species that are present in the second cell but not in the first.

The second $\beta$-diversity considered in this study is based on the turnover component of the Bray-Curtis dissimilarity index \citep{Baselga2013}, calculated between a given pair of distinct cells with the following formula:
\begin{equation}
	\beta^P_{BC-bal} = \frac{min(B_P,C_P)}{A_P+min(B_P,C_P)}
	\label{betabcbal}
\end{equation}
Following the same reasoning, $A_p$ may be seen as the intersection based on the species abundance in two distinct cells $k$ and $l$, and $B_p$ and $C_p$ as the relative complements of each cell \citep{Baselga2013}, which can be formulated as:
\begin{align}
	\phantom{i + j + k}
	& \begin{aligned}
		A_P & = \sum_{p=1}^P min(X_{pk},X_{pl})
	\end{aligned}\\
	&\begin{aligned}
		B_P & = \sum_{p=1}^P X_{pk} - A_P 
	\end{aligned}\\
	&\begin{aligned}
		C_P & = \sum_{p=1}^P X_{pl} - A_P
	\end{aligned}     
\end{align}
where $X_{pk}$ stands for the number of observations of species $p$ in cell $k$.

Here again, both indices were also computed for the spectral clusters, $\beta^S_{SIM}$ and $\beta^S_{BC-bal}$, by replacing $s$ with $p$ and $S$ with $P$ in Equation \ref{betasim} and \ref{betabcbal}, respectively.

All the $\beta$-diversity metrics were computed using the R package \textit{bioregion} \citep{Denelle2025}.

\begin{figure*}[!ht]
	\centering 
    \includegraphics[width=\linewidth]{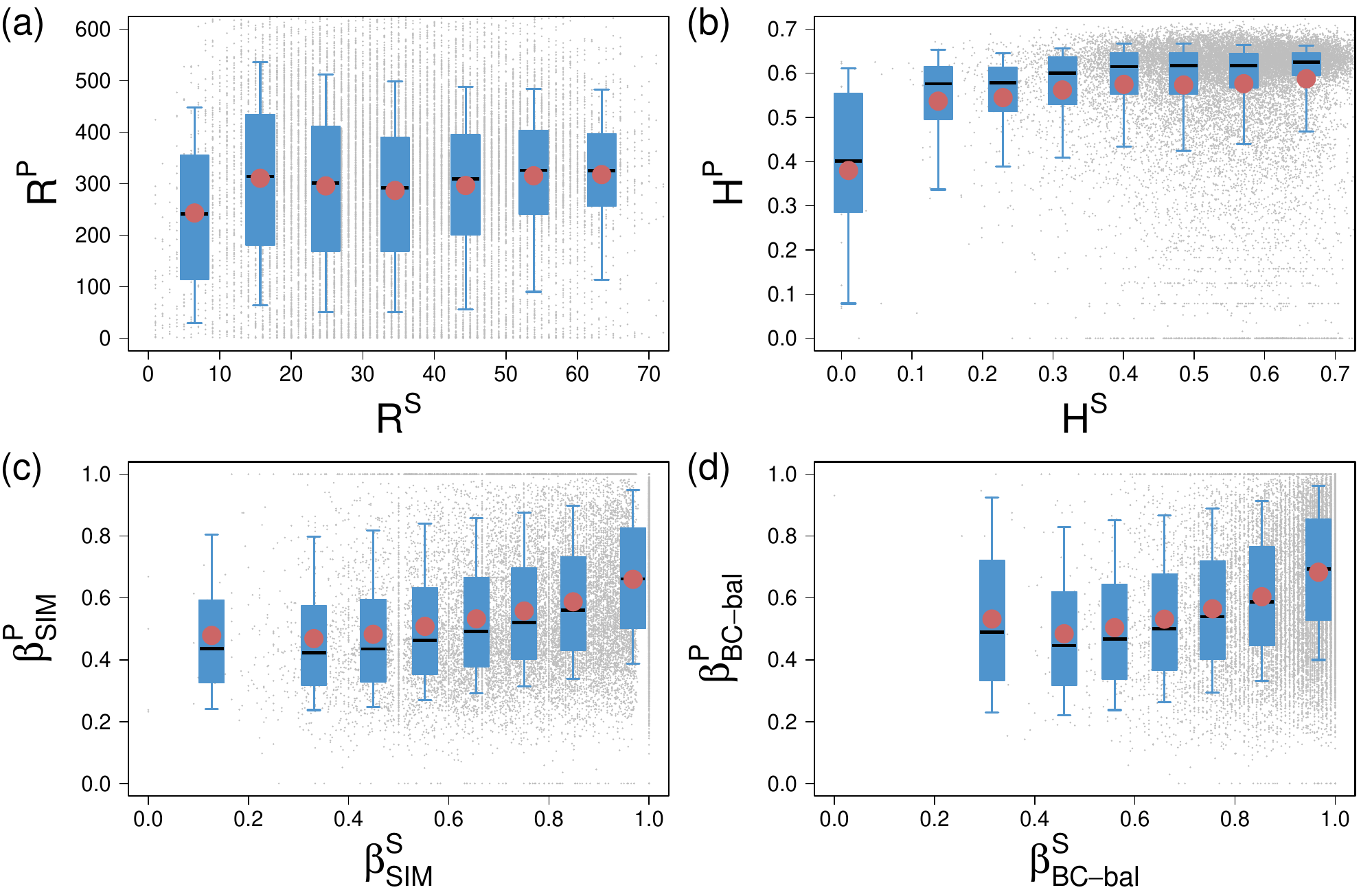}
	\caption{\textbf{Comparison of $\bm{\alpha}$-diversity and $\bm{\beta}$-diversity indices obtained with plant species and spectral cluster distributions.} Comparison of (a) the number of plant species and spectral clusters per cell ($R$), (b) the Shannon diversity index per cell ($H$), (c) the Simpson dissimilarity index per pair of cells ($\beta_{SIM}$) and (d) the turnover component of the Bray-Curtis dissimilarity index per pair of cells ($\beta_{BC-bal}$). Grey points are the scatter plot for each cell ((a) and (b)) or a sample of 20,000 pair of cells ((c) and (d)). The boxplots (D1, Q1, Q2, Q3 and D9) represent the distribution of the y-axis index in different bins of the x-axis index. The red dots represent the average in each bin. Only cells with at least one plant species and one spectral cluster were considered.}
    \label{Fig2}
\end{figure*}

\subsection*{Identification of bioregions}

To understand the importance of spatial context in comparing the two data sources, we needed to compare the metrics described above on a national and regional scale. To do this, we identified meaningful biogeographical regions that capture a common spatial structure between plant species and spectral clusters.

To quantify the intensity of spatial interactions between plant species and spectral clusters, we constructed a bipartite network, where a plant species and a spectral cluster are connected if they are both present in at least one cell. The edge weight represents the weighted spatial overlap between the spatial distributions of a plant species and a spectral cluster. The weighted overlap coefficient used in this study is based on the turnover component of the Bray-Curtis similarity index \citep{Baselga2013}. In order to remove non-significant and null values, only weighted overlap coefficients above a threshold of $0.2$ were considered. More details on the construction of the bipartite network and the identification of thresholds can be found in the Appendix.

We then applied the Louvain algorithm \citep{Blondel2008} to the bipartite network to identify network communities of plant species and spectral clusters that share similar spatial features. Each network community contains both plant species and spectral cluster. A network community can also be assigned to each cell by calculating the proportion of species/clusters belonging to a given community present in that cell. Since the number of plant species is significantly higher than the number of spectral clusters, and since these numbers can be very different from one network community to another, we decided to assign each species or cluster a weight that varies according to the type of species or cluster and its network community. Therefore, the weight of a plant species (or spectral cluster) belonging to a given community is inversely proportional to the number of plant species (or spectral cluster) in that community. Thus, the proportion of species/clusters present in a given cell and belonging to a given network community is equal to the ratio of the sum of the weights of these species/clusters to the total sum of the weights of the species/cluster present in that cell. We then assigned each cell to the network community with the higher proportion of species/clusters in that cell. The set of cells belonging to a given network community is hereafter referred to as a \enquote{bioregion}. More details on the identification of bioregions can be found in the Appendix.

\subsection*{Correlation measures}

We focused on Pearson's correlation coefficient to compare the four diversity metrics obtained with the plant species and spectral cluster distributions. Unless otherwise stated, the correlation measures are based on the grid cells with at least one plant species and one spectral cluster (see \textit{Data filtering} section in the Appendix for more details).

Correlations were measured at the national scale but also within each bioregion. Correlations were also measured between bioregions by calculating the turnover component of $\beta$-diversity between cells of two different bioregions.

\begin{figure*}[!ht]
	\centering 
	\includegraphics[width=16cm]{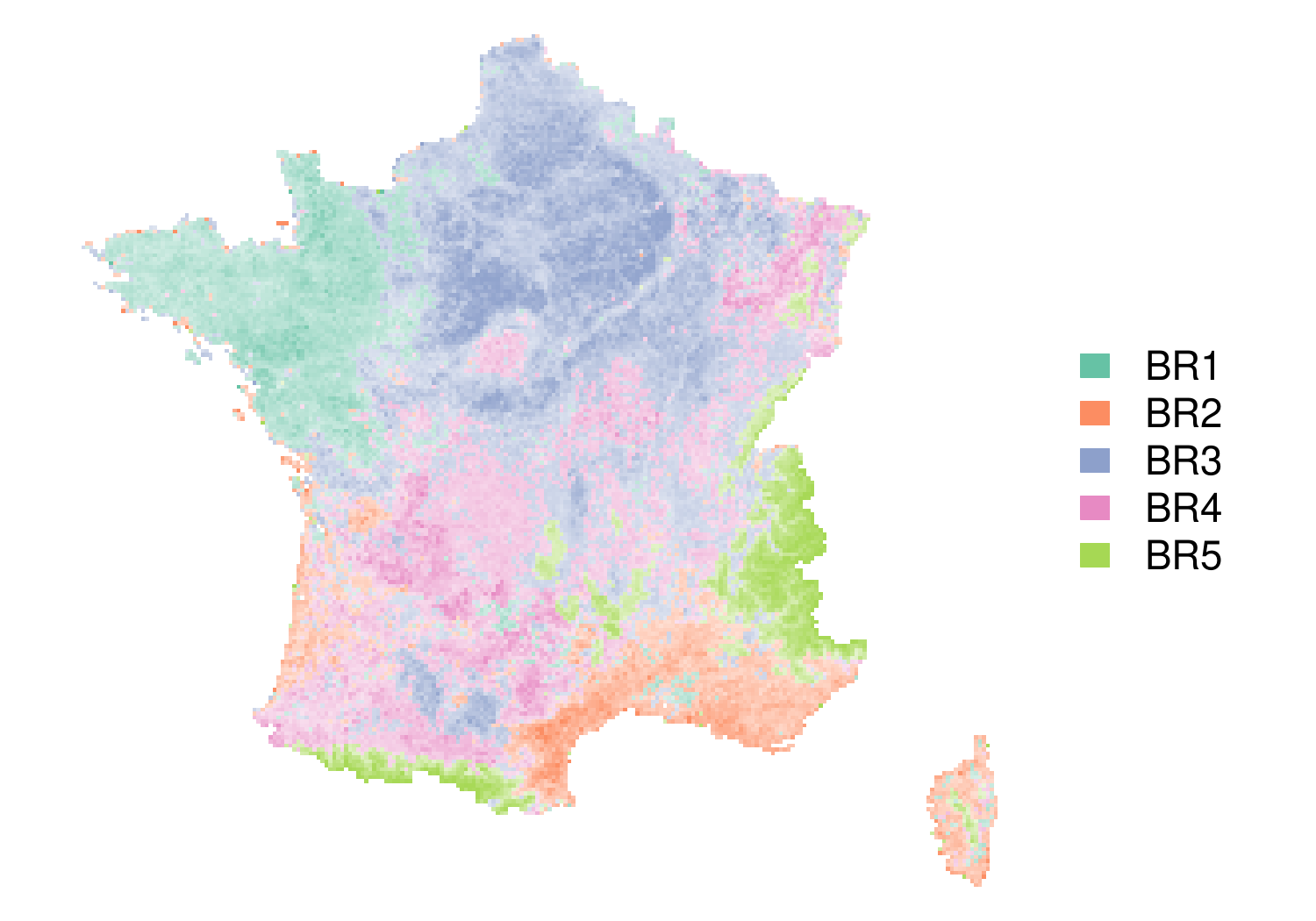}
	\caption{\textbf{Bioregion identification.} Map of France displaying the five bioregions with the colors varying from white to the baseline color of the network community assigned to each cell. The intensity of color depends on the fraction of species/clusters associated with the network community assigned to the cell.}
\label{Fig3}
\end{figure*}

\section*{Results}

\subsection*{Global analysis}

A comparison of $\alpha$-diversity and $\beta$-diversity indices obtained with plant species and spectral clusters was performed (Figure \ref{Fig2}). Figure \ref{Fig2}a shows a scatterplot of $R^S$ and $R^P$ for each pair of cells. The plot shows a weak positive relationship (Pearson's correlation coefficient of 0.037), highlighting little to no biological relationship between plot species richness and spectral richness captured. Results are globally similar when assessing $\alpha$-diversity using the Shannon diversity index, despite a slightly higher correlation (Pearson's correlation of 0.072).

The correlations between the $\beta$-diversity indices displayed in Figure \ref{Fig2}c and \ref{Fig2}d show more robust trends, with a Pearson's correlation of 0.281 for the Simpson dissimilarity index and 0.261 for the turnover component of the Bray-Curtis dissimilarity index. This depicts a general coherence in the ability of the two approaches to detect turnover between cells.

\begin{table}[!h]
	\caption{\textbf{Pearson's correlation coefficient between diversity indices (and the associated 95\% confidence interval in square brackets and significance level in parentheses) obtained with plant species and spectral cluster distributions.} The first column shows the results considering each cell with at least one plant species and one spectral cluster have been considered. The second column corresponds to the correlation coefficient obtained on the subsample of grid cells, filtering out cells with a low number of plant species and/or spectral clusters, or exhibiting a relationship between the number of observations and the number of plant species deviating from the main data trend (more details are available in the \textit{Data filtering} section in the Appendix). Significant correlations are marked with *** (p $<$ 0.001), while non-significant correlations are indicated as \enquote{ns}.}
	\label{Tab1}
	\centering
	\vspace{0.2cm}
	\begin{tabular}{cccc}		
		\hline
		\textbf{Index} & \textbf{Pearson (no filter)} & \textbf{Pearson (filter)}\\	
		\hline					
		\bm{$R$} & 0.037 [0.024,0.05] (***) & 0.012 [-0.003,0.027] (ns)\\
		\bm{$H$} & 0.072 [0.059,0.086] (***) & 0.128 [0.113,0.143] (***)\\
		\bm{$\beta_{SIM}$} & 0.281 [0.281,0.282] (***) & 0.362 [0.362,0.362] (***)\\
		\bm{$\beta_{BC-bal}$} & 0.261 [0.261,0.261] (***) & 0.336 [0.336,0.336] (***)\\		
		\hline          	
	\end{tabular}
\end{table}

In order to understand the influence of various forms of sampling and measurement uncertainty on the results, we also computed the correlation coefficients by discarding cells with a low number of plant species and/or spectral clusters or exhibiting a relationship between the number of observations and the number of plant species deviating from the main data trend (more details available in \textit{Data filtering} section in Appendix). The results obtained with this new sample of 16,551 cells are available in Table \ref{Tab1}. These results confirm the very weak correlation between $R^S$ and $R^P$, which is not significant in this case. However, we observe an increase of Pearson's correlation between $H^S$ and $H^P$ (from 0.072 to 0.128), $\beta^S_{SIM}$ and $\beta^P_{SIM}$ (from 0.281 to 0.362), and, $\beta^S_{BC-bal}$  and $\beta^P_{BC-bal}$ (from 0.261 to 0.336).

\begin{figure*}[!ht]
	\centering 
	\includegraphics[width=\linewidth]{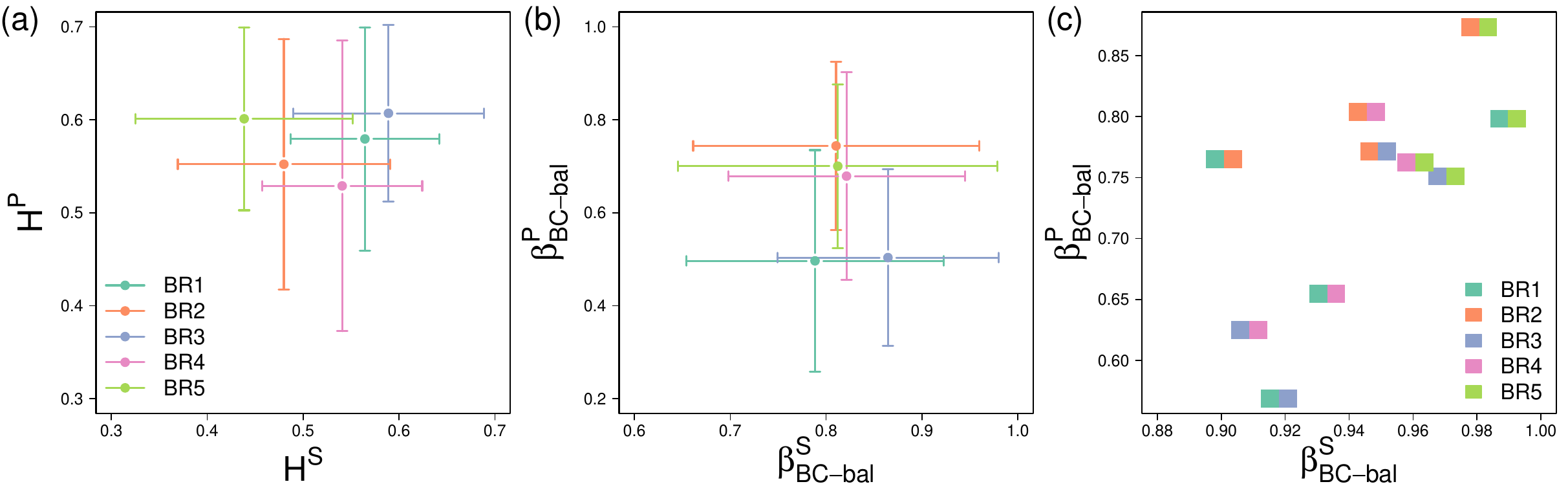}
	\caption{\textbf{Comparison of $\bm{\alpha}$-diversity and $\bm{\beta}$-diversity indices obtained with plant species and spectral cluster distributions within and between bioregions.} (a) Average Shannon diversity index per cell ($H$) within bioregion (and its associated standard deviations). (b) Average turnover component of the Bray-Curtis dissimilarity index per pair of cells ($\beta_{BC-bal}$) within bioregion (and its associated standard deviations). (c) Average turnover component of the Bray-Curtis dissimilarity index per pair of cells ($\beta_{BC-bal}$) between bioregions. Standard deviations are available in Table S2 in the Appendix.}
\label{Fig4}
\end{figure*}

\subsection*{Bioregional analysis}

We applied the Louvain algorithm \citep{Blondel2008} and identified five plant-spectral communities in the bipartite network. Details regarding the intensity of the spatial interaction (i.e., the weight of the bipartite network) between plant species and spectral clusters according to the network community are available in the \textit{Identification of bioregions} section in the Appendix. The percentages of plant species, spectral clusters and grid cells by network community are available in Table S1 in Appendix.

By assigning a network community to each grid cell, we identified five bioregions reflecting the biogeographical structure of France based on plant species and spectral cluster distributions (Figure \ref{Fig3}). The biogeographical structure shown is consistent with several empirical studies, such as that of GRECO\footnote{\url{https://inventaire-forestier.ign.fr/spip.php?article773}} on the forests of France \citep{Bontemps2019}. The first bioregion (BR1) covers much of the westernmost part of France and adjacent regions. This zone is characterised by a marked oceanic climate and a substratum dominated by the Hercynian crystalline rocks of the Armorican massif. The second bioregion (BR2) corresponds mainly to France's Mediterranean zone and includes the rocky west's coastal vegetation. The third bioregion (BR3) mainly covers the plains of northern France. Under the oceanic influence, these areas include large agricultural plains mainly occupied by neutral-basic soils and Quaternary sedimentary deposits. The fourth bioregion (BR4) is a relatively heterogeneous group distributed throughout the country. It includes large areas of medium-sized mountains (in the Massif Central and the Vosges) and various lowland areas. The fifth bioregion (BR5) covers high mountain areas, mainly in the Alps, the Jura, and the southern Pyrenees, with more peripheral zones in Corsica, the Massif Central, and the Vosges mountain ranges. The substratum is largely heterogeneous, but the bioregion is characterized by a common topography and vegetation structure from the mountain to the alpine zone.

We plot in Figure \ref{Fig4} the average values of two $\alpha$-diversity and $\beta$-diversity ($H$ and $\beta_{BC-bal}$) obtained within each bioregion and also between bioregions for the $\beta$-diversity. Figure \ref{Fig4}a shows the relationship between plant and spectral $\alpha$-diversity by bioregion. Globally, the $H$ index behaves similarly for plant species and spectral clusters within the same bioregion. Only BR5 (mountain region), rich in plant species, deviates from the overall pattern, with plant $\alpha$-diversity higher than spectral $\alpha$-diversity.

Figure \ref{Fig4}b shows the relationship between the Bray-Curtis turnover estimated from flora and spectral data within each bioregion. The $\beta$-diversity index obtained with spectral clusters is spread over a smaller range of values (0.78 to 0.86) than that obtained with plant species (0.5 to 0.78). The plant data divide the bioregions into two groups: BR2/BR5/BR4 with high values for the plant index, reflecting greater spatial heterogeneity of plant communities, and BR1/BR3 with lower values. Bioregion BR3 has the highest value for the spectral clusters but one of the lowest for the plant species. This can be partially explained by the prevalence of natural and semi-natural habitats in the first group (Mediterranean, middle and high mountains) in preference to an agricultural matrix, suggesting a certain biotic homogenization despite the marked landscape structure.

Figure \ref{Fig4}c represents the average Bray-Curtis turnover between each pair of bioregions estimated with plant species and spectral cluster. The two approaches have a positive relationship, with the highest values for the Mediterranean (BR2) and mountain (BR5) zones, which are very different from the other regions for both approaches. Only the BR1-BR2 comparison shows a higher dissimilarity for plant species than for spectral cluster, probably indicating a very different species pool (medio-European versus Mediterranean) but a less dissimilar vegetation structure. Conversely, the lowest values between BR1, BR3, and BR4 reflect the greater similarity between these entities, with the two lowland bioregions with an agricultural matrix (BR1/BR3) showing the greatest similarity.

\begin{figure*}[!ht]
	\centering 
	\includegraphics[width=\linewidth]{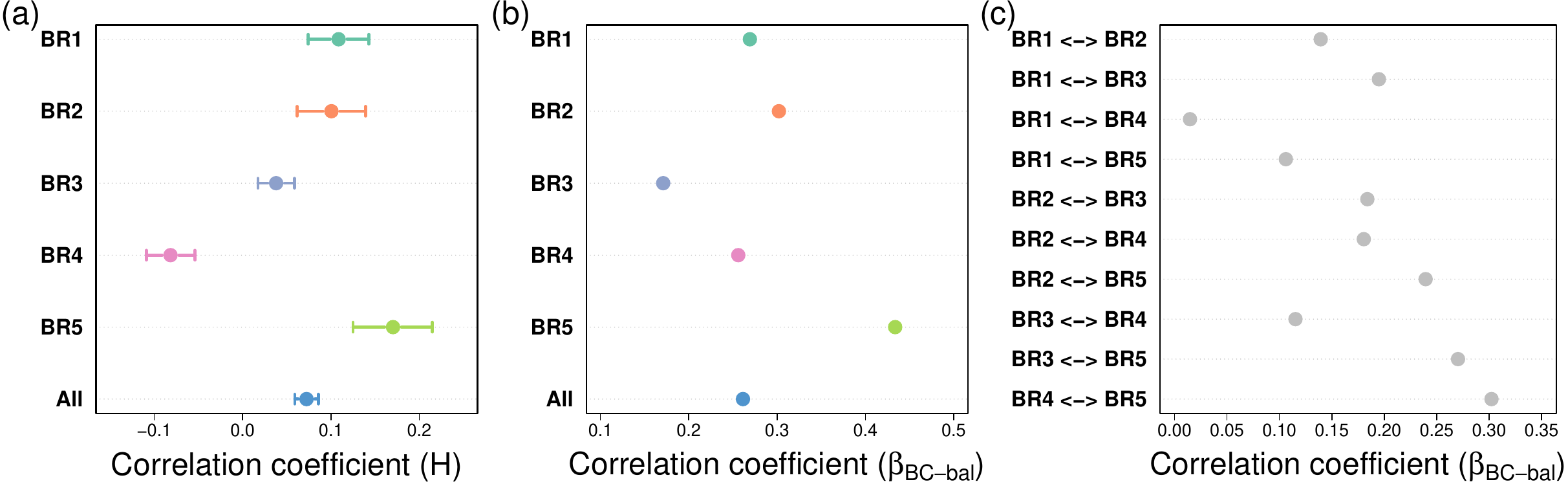}
	\caption{\textbf{Correlation between $\bm{\alpha}$-diversity and $\bm{\beta}$-diversity indices obtained with plant species and spectral cluster distributions within and between bioregions.} (a) Pearson's correlation coefficients between Shannon diversity indices obtained with plant species and spectral cluster for each bioregion and for the whole study area. (b) Pearson's correlation coefficients between the turnover component of the Bray-Curtis dissimilarity index obtained with plant species and spectral cluster for each bioregion and for the whole study area. (c) Pearson's correlation coefficients between the turnover component of the Bray-Curtis dissimilarity index obtained with plant species and spectral cluster between bioregions. Error bars indicating the 95\% confidence interval are too small to be seen for the turnover component of the Bray-Curtis dissimilarity index. In all cases, the correlation coefficient is significantly different from 0, with p-values consistently below 0.001. See Table S3 in Appendix for more details.}
\label{Fig5}
\end{figure*}

Figure \ref{Fig5} shows Pearson's correlation coefficients (and their 95\% confidence intervals) by bioregion. Figure \ref{Fig5}a shows the correlation coefficient between Shannon diversity indices obtained with plant species and spectral cluster in the whole study area (i.e., \enquote{All}) and within bioregions. The overall correlation is globally weak, with values per bioregion ranging from -0.1 to 0.17. BR4 shows a weaker correlation and BR5 a stronger one, reflecting a more robust relationship between $\alpha$-diversity estimated with plant species and spectral cluster. Thus, bioregions have no substantial structuring effect on the correlation between $\alpha$-diversity estimated with plant and spectral information.

Figure \ref{Fig5}b displays the Pearson correlation coefficient between plant species and spectral clusters obtained with the turnover component of the Bray-Curtis dissimilarity index for the whole study area (i.e., \enquote{All}) and within bioregions (Figure \ref{Fig5}b). We observed correlation coefficients close to the average for the BR1/BR2/BR4 bioregions (~0.3) and slightly lower for BR3 (0.17). Only BR4 (mountainous zone) exhibits a stronger correlation, highlighting a higher plant turnover when spectral turnover is detected.

Finally, Figure \ref{Fig5}c displays the Pearson correlation coefficient between plant species and spectral clusters obtained with the turnover component of the Bray-Curtis dissimilarity index between each bioregion (Figure \ref{Fig5}c). These coefficients are mostly centered around the value of 0.15, with extremes ranging from 0.01 (BR1/BR5) to 0.3 (BR4/BR5). These high values suggest that a high turnover of spectral clusters translates into a high turnover of plant species.

\section*{Discussion}

Analyzing the spatial structure of diversity patterns is key to understanding and managing biodiversity \citep{Wang2019,Carvalho2021,Geppert2020,Hallmann2021, Overcast2021}. These approaches can benefit from new technologies. In this study, we compared and combined field data describing plant communities in mainland France with diversity metrics computed from phenological information of vegetation derived from moderate spatial resolution RS data to describe the structure of the French biogeographical structure using a network approach. We analyzed the consistency of the two sources of information to estimate $\alpha$-diversity and $\beta$-diversity within and between bioregions. We found a coherent biogeographic structure despite a heterogeneous floristic dataset. Through this discussion, we will address how RS analysis can prove complementary to in situ data to assess different facets of the biogeographic diversity of plant assemblages. We will also discuss the limitations of the two approaches and how future methodological developments will allow them to be better combined.

\subsection*{Comparison of $\alpha-$diversity}

The first step in our study was to analyze the consistency between field diversity metrics and remotely sensed diversity metrics derived from moderate resolution vegetation phenology, for estimating diversity within each $5 \times 5$ km$^2$ grid cell. The RS approach is based on virtual \enquote{spectral clusters}, defined by a set of spectro-temporal features corresponding to the monthly NDVI time-series over one year. In contrast, the field approach is based on a list of species derived from the synthesis of opportunistic inventories. The two approaches showed a lack of correlation for taxonomic richness (RS and plant) and a very weak correlation for Shannon’s diversity index. This was independent of the detected bioregions, with only the high mountains showing a slightly stronger correlation coefficient.

This lack of consistency is not surprising given the spatial heterogeneity of floristic inventories (Figure S1b in the Appendix) and the spatial patterns of spectral cluster richness. Indeed, the RS approach detects the highest levels of richness in the agricultural areas of north-central France and the lowest levels of richness in the high and medium mountain areas and the Mediterranean region. This contrasts completely with the empirically known patterns of plant biodiversity \citep{Biurrun2021} and primarily reflects the limitations of the spectral cluster approach applied to moderate spatial resolution imagery \citep{Rocchini2014}. It is likely that the local spatial heterogeneity induced by various components within agricultural systems (such as neighboring fields growing different crops with diverse agricultural practices, meadows, and edges) surpasses that found in natural and semi-natural environments like alpine grasslands and Mediterranean shrubland. Marked differences in the phenology of specific crops may also have augmented spectral diversity. Therefore, it is reasonable to assume that diversity in surface phenology derived from moderate spatial resolution NDVI is not directly comparable to biological species diversity. The estimation of the former relies on local contingencies that are independent of plant biodiversity.

\subsection*{Comparison of $\beta-$diversity}

During a second step, the analysis of $\beta$-diversity using taxonomic and RS approaches showed greater consistency between the two approaches.The analysis of $\beta$-diversity patterns emphasized a turnover between pairs of grid cells, thereby incorporating unit identity (both spectral and taxonomic) to measure shifts within communities. This approach revealed a notable level of global consistency, with results consistently observed across various bioregions, indicating the absence of significant local regional outliers in the overall findings. Nevertheless, it’s worth noting that this signal exhibited slight elevation in high mountain regions and a decrease in lowland arable areas. One possible interpretation is that turnover in the mountains is structured by a marked change in natural habitats, the taxonomic composition of which is largely specific. This change in natural habitat is detected by an equivalent change in spectral clusters and floristic assemblages. Alternatively, the spectral clusters will detect different crops (particularly due to their specific phenology), which, oppositely, form a globally homogeneous habitat with poor flora (i.e., ruderal weeds and generalist species).

In addition to these methods, comparing the beta diversity of grid cells from distinct bioregions provides valuable insights into the inferred degree of ecological differentiation between two spatial entities. Taxonomic and spectral approaches detect the same strong turnover between the Mediterranean (BR2), high mountain (BR5) and mid-mountain (BR4) regions. These zones differ greatly in their species assemblages, with the most marked transition between the hot, dry Mediterranean zone (BR3) and the cold high mountain zone (BR5). In the latter, the flora is under a Eurobasian influence, starkly contrasting with Mediterranean biodiversity. This is also reflected in the marked contrasts between the habitats that dominate the landscape: among open habitats, Alpine perennial grassland differs in every way from open shrubland with stronger contribution of bare soil; similarly, deciduous and mixed coniferous/hardwood forests exhibit different spectral signals from Mediterranean evergreen forests, ultimately reflected in a marked turnover of spectral clusters. In contrast, the agricultural plain zones (BR3) and the oceanic zone on an acid substratum (BR1) show the greatest floristic similarity and a strong phenological similarity. This reflects a shared biogeographical influence, with widely distributed temperate plant species and relatively similar spectral clusters, suggesting similar agricultural landscape components (e.g., fields and grasslands) and deciduous forests.

Therefore, while spectral approaches have shown limitations in assessing patterns of floristic richness, they are generally consistent in evaluating spatial turnover.

\subsection*{Combining remote sensing and field data to identify bioregions}

The approach proposed in this article allows for identifying bioregions based on a minimum common spatial structure between plant species and spectral clusters (Figure \ref{Fig3}). The overall result is consistent with the bioregionalization proposed by several authors \citep{Lenormand2019,Bontemps2019}. The high mountain areas of France, in particular the Alps and the Pyrenees, are well-defined. This vegetation type is also found on the summits of the lower massifs, including the Vosges, the Jura, the Massif Central and Corsica. The common floristic influences and vegetation structure between mixed coniferous and deciduous forests, alpine meadows, and high rocky environments create uniformity in these areas. The presence of snow for part of the year reinforces this similarity. The delineation of the Mediterranean zone is also consistent with recent work \citep{Lenormand2019}. There is a link with the Atlantic coastal areas, which, although highly contrasted in species assemblages, could contain a relatively similar structuring of coastal habitats (sandy dune or rocky coastline).

Therefore, coupling RS and in situ data could significantly contribute to fill gaps in biodiversity monitoring \citep{Vihervaara2017}. Firstly, it is generally accepted that the quality of taxonomic field data will condition the possibility of carrying out relevant bioregionalization \citep{Kreft2010}. In particular, the spatial accuracy and the sampling effort \citep{Giraudo2018,Ferrier2002} will condition the final resolution of the work and define the grain at which the spatial structure of bioregions can be assessed. Despite the constant increase in the amount of naturalist data, the latter remain disparate in many areas of the globe, and the absence of structured sampling limits the observation pressure on certain taxonomic groups \citep{August2020,Reichman2011,Isaac2014}. Secondly, it is important to remember that bioregionalization approaches are almost always carried out at a given time: collecting field data on large spatial scales is far too costly to be replicated over time to assess bioregions' temporal dynamics. This approach would enable us to dynamically understand bioregions, which could be relevant in transition zones between bioregions in the context of climate change and related migration \citep{Lenormand2019}.

\subsection*{Limitations of the study}

The definition of a bioregion depends on the quality of the dataset available. Here, floristic data were aggregated at a national level based on floristic inventories conducted by  botanical conservatories assigned to different administrative regions. This dataset had significant gaps, especially in areas which were not assigned to local  conservatories (e.g., in the northeast) (see Figure S1b in the supplementary information). At the same time, sampling pressure within our dataset is highly variable between institutions (Figure S1a in the supplementary information), suggesting a highly variable state of knowledge within each region. These biases have affected the delineation of bioregions, although mitigated by the RS information. In the future, the major challenge will be understanding the relative contribution of the two approaches at different locations in the network to partition the importance of taxonomic and RS data sources on the final result.

In addition to these sampling issues, the  taxonomic resolution also has potentially important implications for this comparison. The data for this paper have been aggregated at a certain level according to a common reference system to standardize the taxonomic level. The main drawback of this approach is that it flattens some of the knowledge by removing the finesse of identifying population groups, often defined at the subspecies level, which could prove informative for understanding certain biogeographical structures. In addition, taxonomic units are considered at the same level, with no interdependence to their evolutionary similarity. A potential development of these methods could be to become taxonomically explicit by incorporating a distance between taxa to refine the definition of bioregions and reveal a deeper evolutionary structure \citep{Cavender2017,Schweiger2018}.

Finally, it should be noted that the statistically classified individual entities derived from RS data (i.e., pixels) are independent within the analysis. A possible improvement would be to make them spatially explicit, allowing the inference of a bioregion to under-sampled pixels based on spectral clusters coupled with a biogeographic signal inferred from the spatial proximity of other pixels.  

\subsection*{Perspectives}

Identifying bioregions is essential for managing ecosystems and developing conservation policies \citep{Woolley2020}. Creating coherent geographical clusters based on shared biodiversity is crucial for better understanding a region’s ecological and evolutionary structure. These tools are particularly useful for ecological restoration and the establishment of protected areas \citep{Smith2004}.

Therefore, it may be worthwhile to develop these approaches in areas where taxonomic knowledge is limited and for groups of species that need better documentation. Improving the definition of bioregions, coupled with geospatial data, proves relevant and opens the way to developing an approach that integrates ecologically, spatially, and phylogenetically explicit advances.

\subsection*{Data and code availability statement} 

Data and code used in this paper are available online (\url{https://zenodo.org/records/14973870}).

\vspace*{0.5cm}
\section*{Acknowledgments}

The authors would like to thank the Federation of the National Botanical Conservatories (CBN) and the French Biodiversity Agency (OFB) for providing species data from the SIFlore database. A special thank goes to Olivier Argagnon, from CBN Med, for useful discussions. The authors were funded by their salaries as French public servants. The work of JBF was also supported by the Agence Nationale de la Recherche (ANR, France) through the Young Researchers project BioCop (ANR-17-CE32-0001). 

\bibliographystyle{myapalike}
\bibliography{MS}

\onecolumngrid
\vspace*{2cm}
\newpage
\onecolumngrid

\makeatletter
\renewcommand{\fnum@figure}{\sf\textbf{\figurename~\textbf{S}\textbf{\thefigure}}}
\renewcommand{\fnum@table}{\sf\textbf{\tablename~\textbf{S}\textbf{\thetable}}}
\makeatother

\setcounter{figure}{0}
\setcounter{table}{0}
\setcounter{equation}{0}

\newpage
\clearpage
\newpage
\section*{Appendix}

\subsection*{Data description}

\begin{figure}[!h]
	\centering 
	\includegraphics[width=14.5cm]{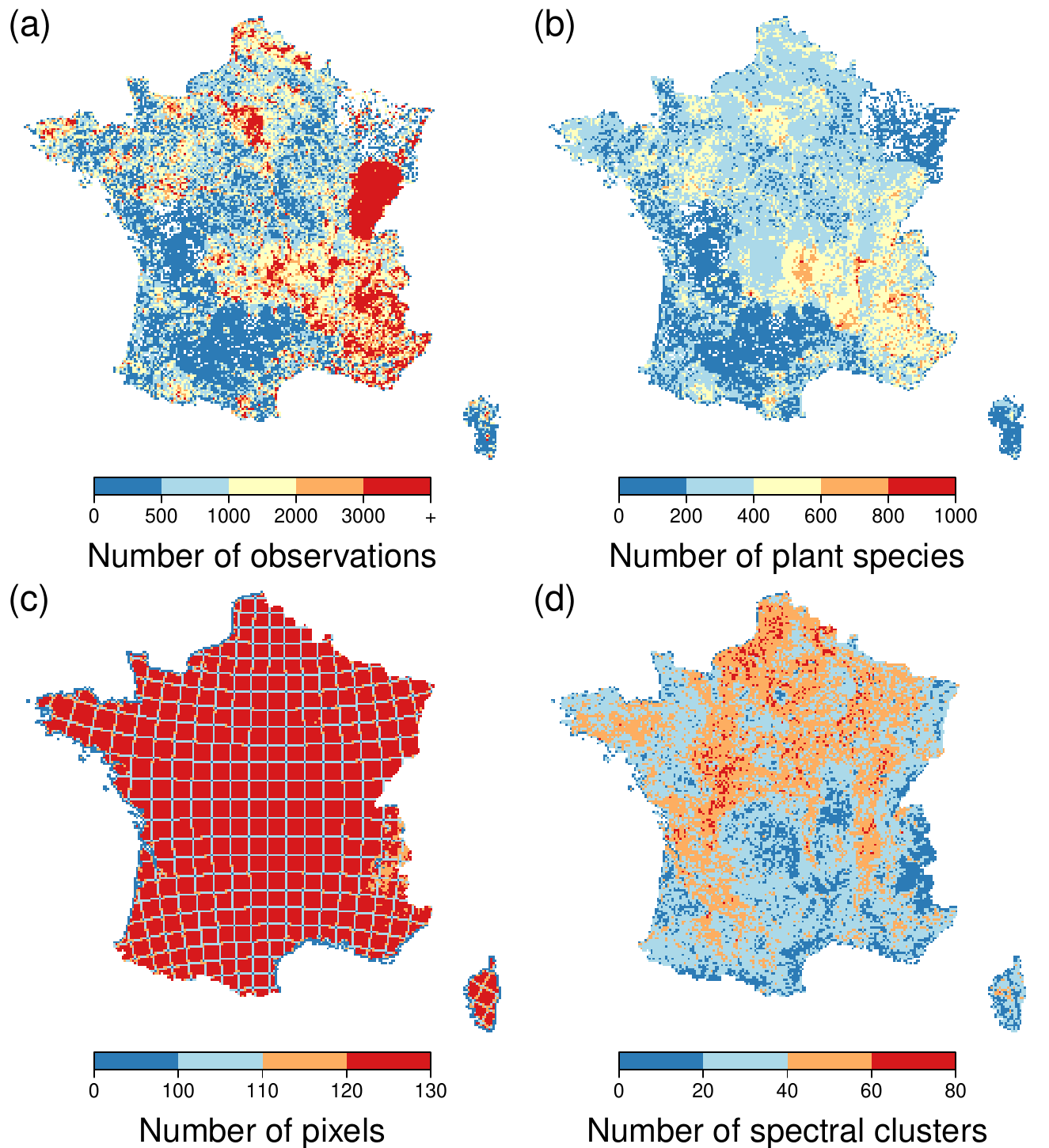}
	\caption{\textbf{Data description.} Distributions of the number of botanic observations (a), number of plant species (b), number of pixels (c) and number of spectral clusters (d) per $5 \times 5\mbox{ km}^2$ grid cell.}
\label{FigS1}
\end{figure}

\newpage
\clearpage
\newpage
\subsection*{Data filtering}

In this study, the map of France has been divided into $N=23,060$ square cells. However, some cells have very few observations for the plant species or pixels for the spectral clusters. First, we only considered cells with at least one observation and one pixel representing 93\% of the cells. Unless otherwise specified, the correlation measures presented in the paper are based on these grid cells.

Then, two consecutive filters have been considered. In order to filter out cells with a number of species/clusters lower than the number of observations/pixels 
we discarded all the cells with less than 200 observations or 20 pixels as can be observed in Figure S2. We obtained a sample of $18,192$ cells (79\% of the original sample). 

\begin{figure}[!h]
	\centering 
	\includegraphics[width=14cm]{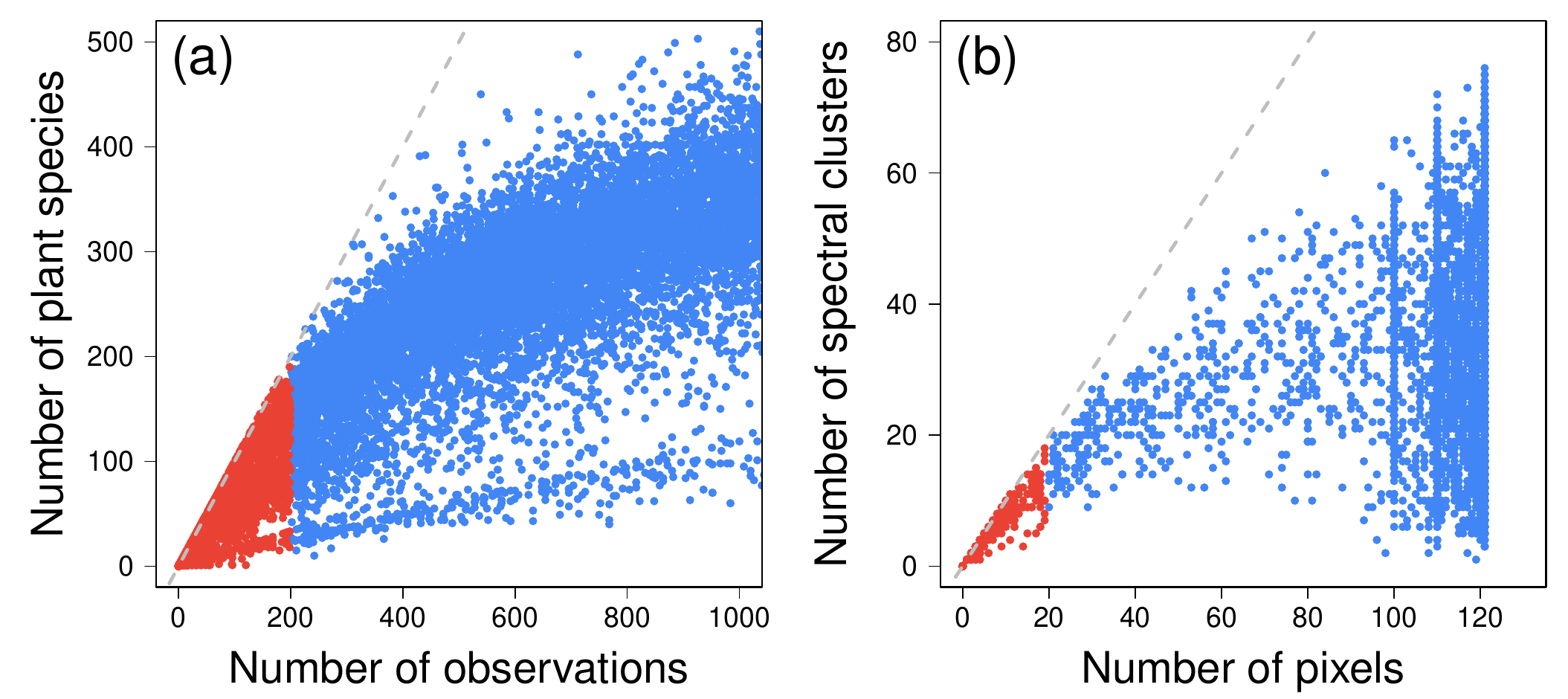}
	\caption{\textbf{Identification of the first filter thresholds.} (a) Number of plant species as a function of the number of observations. Each point represents a cell. The red points represents the cells exhibiting a number of observations lower than 200. (b) Number of spectral clusters as a function of the number of pixels. Each point represents a cell. The red points represents the cells exhibiting a number of pixels lower than 20.}
\label{FigS2}
\end{figure}

We finally filtered out $1,610$ additional cells by applying a second filter to discard cells exhibiting a relationship between the number of observations and the number of plant species deviating from the main data trend (based on Equation S\ref{f2} that can be visualized in Figure S3).  
\begin{equation}
	R^p >´ 1.5 \cdot O^{1/1.5}
	\label{f2}
\end{equation}
At the end of the process we obtained a sample of $16,551$ cells (72\% of the original sample).

\begin{figure}[!h]
	\centering 
	\includegraphics[width=10cm]{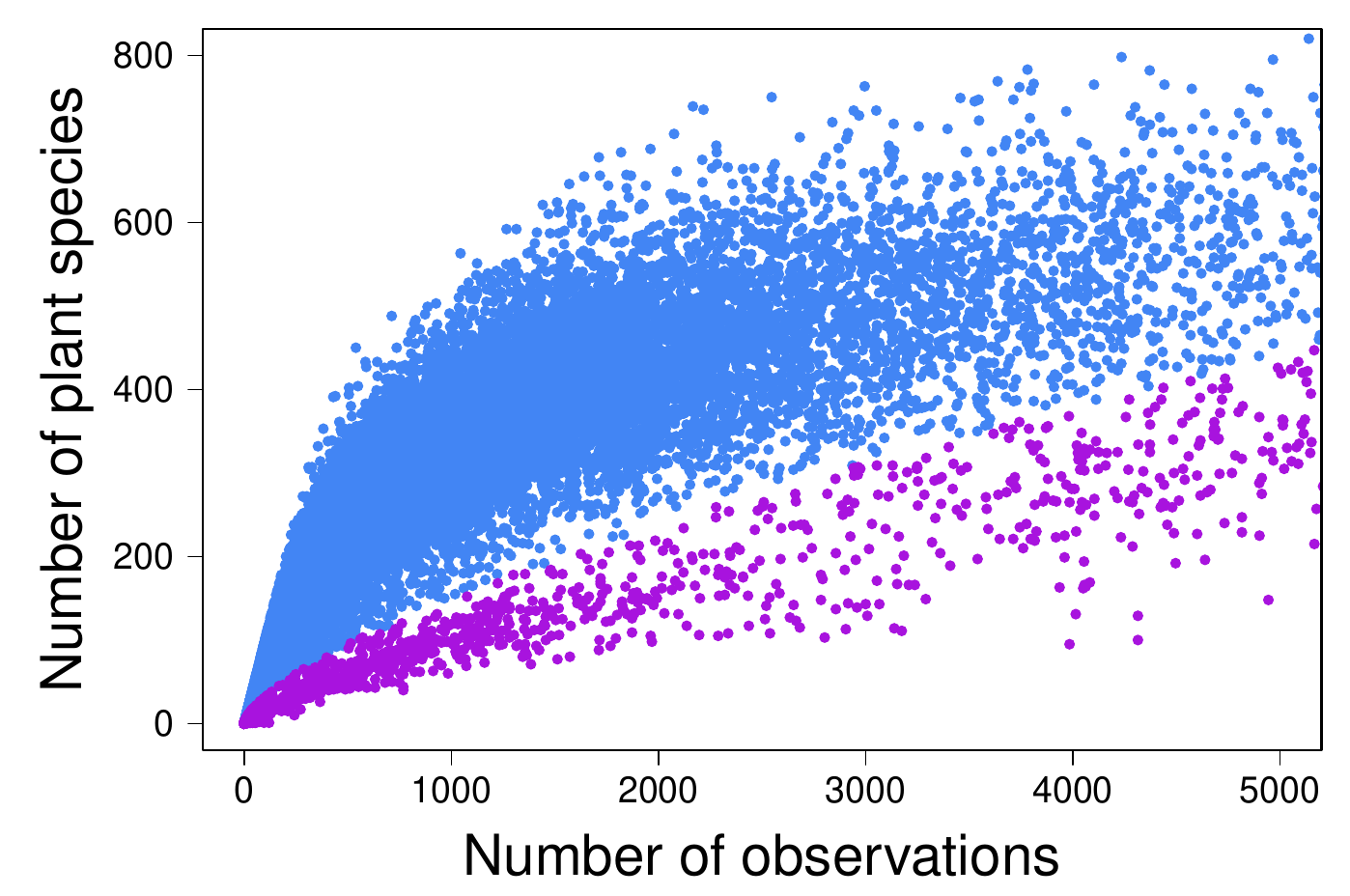}
	\caption{\textbf{Identification of the second filter relationship.} Number of plant species as a function of the number of observations. Each point represents a cell. The purple points represents the cells that did not pass through the second filter (Equation S\ref{f2}).}
	\label{FigS3}
\end{figure}

\begin{figure}[!h]
	\centering 
	\includegraphics[width=13cm]{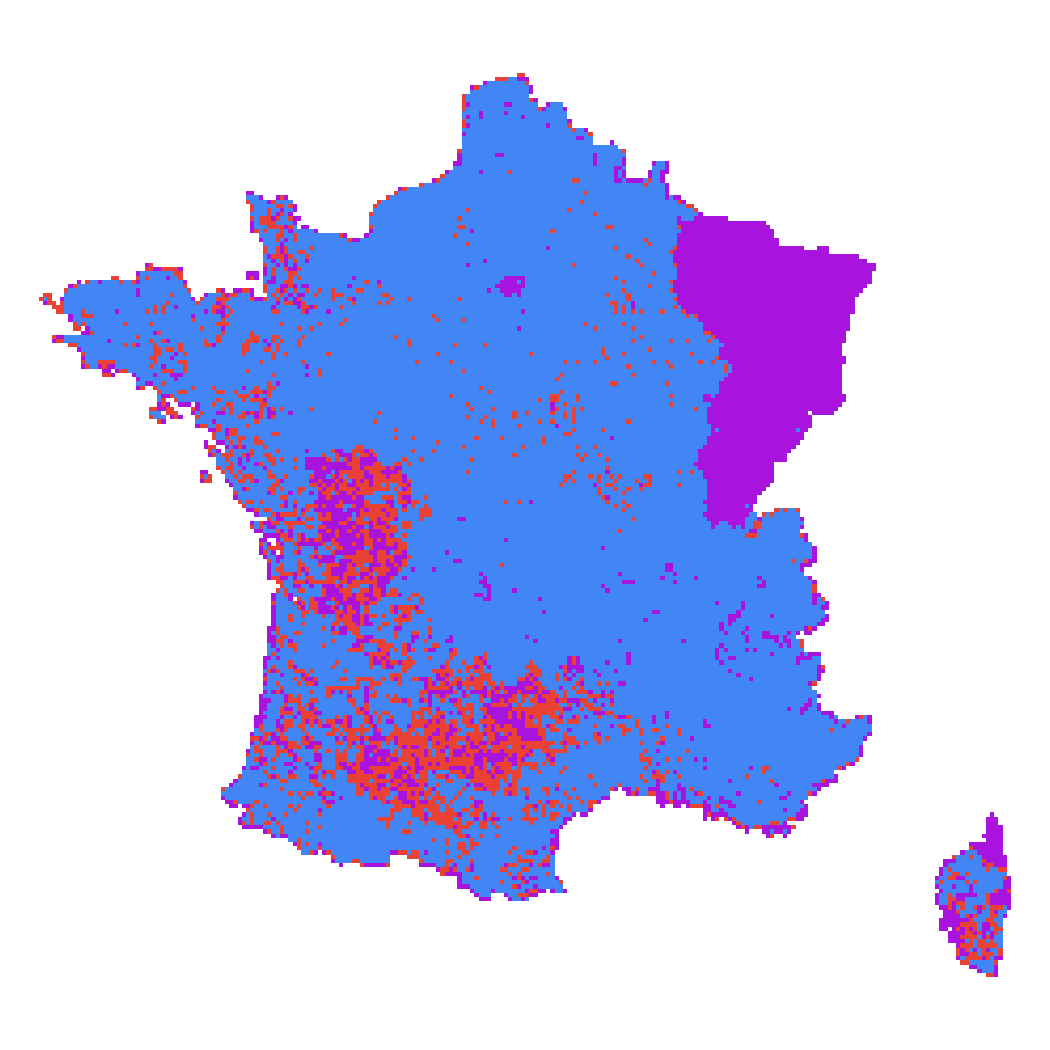}
	\caption{\textbf{Map of the cells colored according to the data filtering.} Red if the cell did not pass through the first filter, purple if the cell did not pass through the second filter and blue if the cell passed through both filters.}
	\label{FigS4}
\end{figure}

\newpage
\clearpage
\newpage
\subsection*{Bipartite network construction}

To quantify the intensity of spatial interactions between plant species and spectral clusters, we built a bipartite network where a plant species $p$ is connected to a spectral cluster $s$ if they are both present in at least one cell. The edge weight $W_{ps}$ represents the weighted spatial overlap between the spatial distributions of the plant species $p$ and the spectral cluster $s$. The weighted overlap coefficient used in this study (Equation \ref{Wps}) is based on the turnover component of the Bray-Curtis similarity index \citep{Baselga2013}. 
\begin{equation}
	W_{ps} = \frac{A_{ps}}{A_{ps}+min(T_p-A_{ps},T_s-A_{ps})}
	\label{Wps}
\end{equation}
where $A_{ps}$ is the number of observations in common between species $p$ and cluster $s$ in grid cells where both are present. $T_p$ and $T_s$ represent the total number of observations of species $p$ and cluster $s$ in the study area, respectively. In order to remove non-significant and null values we only consider weighted overlap coefficients higher than a threshold equal to $0.2$. 

We give here more details on the weighted overlap coefficient computation. Let $X_{pk}$ and $X_{sk}$ be the number of observations of species $p$ and cluster $s$ in cell $k$ (ranging between 1 and $N=23,060$). $A_{ps}$, $T_p$ and $T_s$ are given by the following equations,
\begin{align}
	\phantom{i + j + k}
	& \begin{aligned}
		A_{ps} & = \sum_{k=1}^N min(X_{pk},X_{sk})
	\end{aligned}\\
	&\begin{aligned}
		T_p & = \sum_{k=1}^N X_{pk}
	\end{aligned}\\
	&\begin{aligned}
		T_s & = \sum_{k=1}^N X_{sk}
	\end{aligned}     
\end{align}
Figure S5 shows the probability density function of the bipartite network's edge weights.

\begin{figure}[!h]
	\centering 
	\includegraphics[width=12cm]{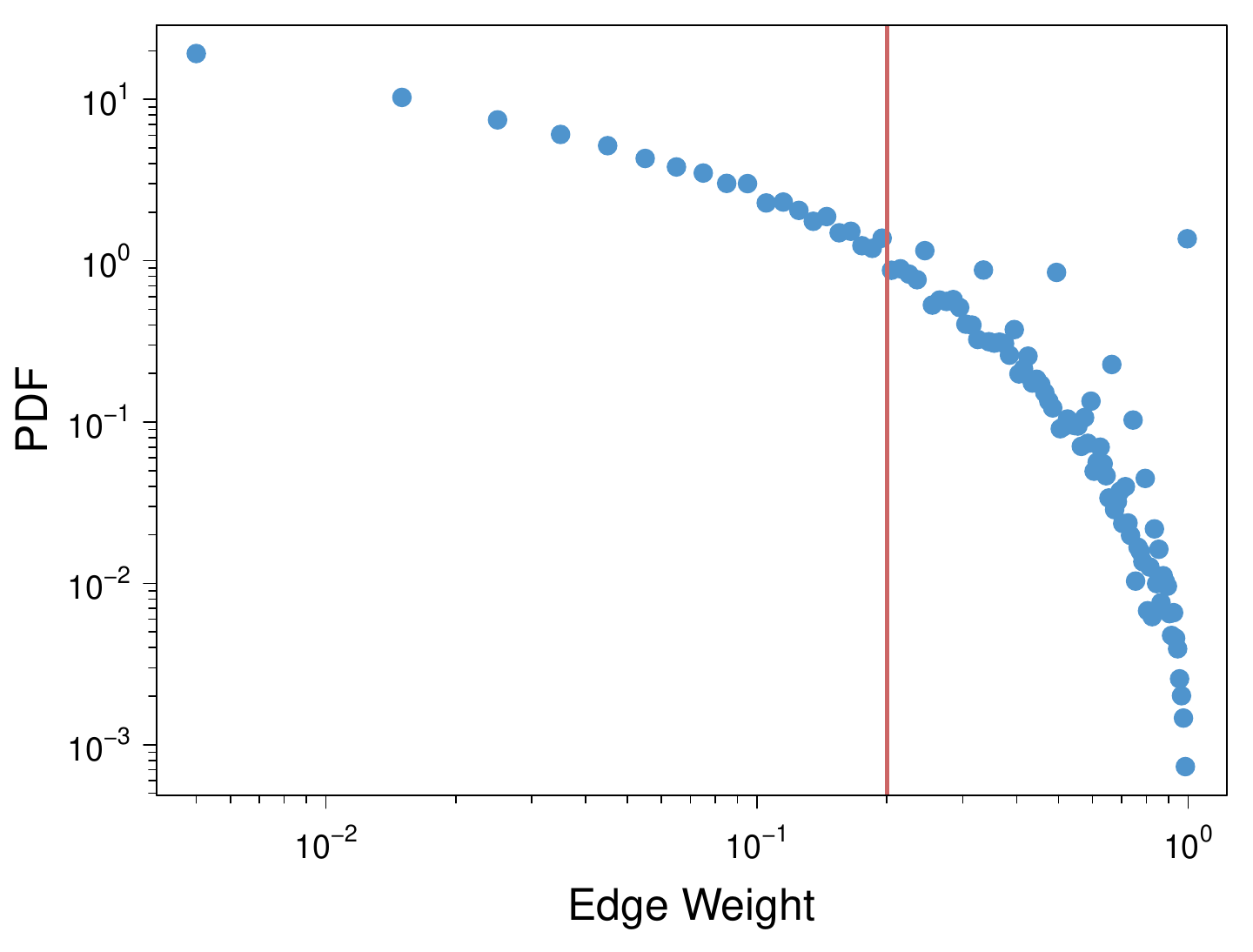}
	\caption{\textbf{Probability density function of the bipartite network's edge weights.} The red line represents the threshold $0.2$ used to prune the network.}
	\label{FigS5}
\end{figure}

\newpage
\clearpage
\newpage
\subsection*{Identification of bioregions}

\begin{figure}[!h]
	\centering 
	\includegraphics[width=11cm]{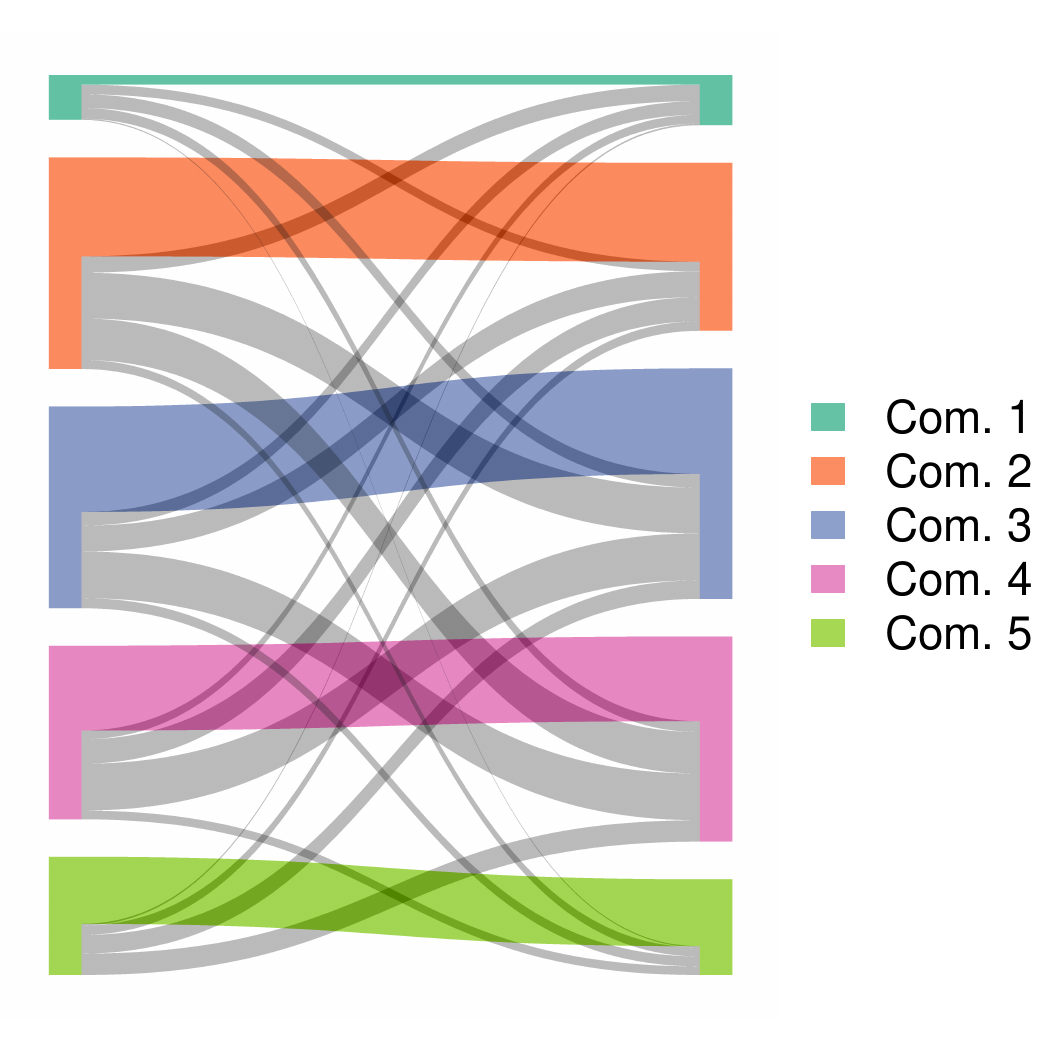}
	\caption{\textbf{Network community detection.} Sankey diagram summarizing the distribution of spatial interactions between plant species and spectral clusters. Each color represents a network community and the width of the line is proportional to the sum of weighted overlap coefficient $W_{ps}$ between plant species (left) and spectral cluster (right) belonging to the same or different network communities.}
\label{FigS6}
\end{figure}

\begin{table}[!h]
	\caption{\textbf{Number (and percentage) of plant species, spectral clusters and grid cells by network community.}}
	\label{TabS1}
	\centering
	\vspace{0.2cm}
	\begin{tabular}{cccc}		
		\hline
		\\[-0.9em]
		\textbf{Community} & \textbf{Plant} & \textbf{Spectral} & \textbf{Spatial} \\ 		
		\hline	
		
		- & 303 (4.56\%) & 0 (0\%) & 183 (0.79\%)\\
		1 & 285 (4.29\%) & 19 (7.6\%) & 3409 (14.78\%)\\
		2 & 2370 (35.64\%) & 38 (15.2\%) & 2718 (11.79\%)\\
		3 & 705 (10.6\%) & 102 (40.8\%) & 9388 (40.71\%)\\
		4 & 1318 (19.82\%) & 62 (24.8\%) & 5458 (23.67\%)\\
		5 & 1669 (25.1\%) & 29 (11.6\%) & 1904 (8.26\%)\\
		
		\hline          	
	\end{tabular}
\end{table}

\newpage
\clearpage
\newpage
\subsection*{Supplementary Tables}

\begin{table}[!h]
	\caption{\textbf{Average turnover component of the Bray-Curtis dissimilarity index per pair of cells  between bioregion (and its associated standard deviations) obtained with plant species and spectral clusters.}}
	\label{TabS2}
	\centering
	\vspace{0.2cm}
	\begin{tabular}{ccc}		
		\hline
		\\[-0.9em]
		\textbf{Relationship} & \bm{$\beta_{BC-bal}$} \textbf{(Plant)} & \bm{$\beta_{BC-bal}$} \textbf{(Spectral)} \\ 		
		\hline	
		
		BR1 $\longleftrightarrow$ BR2 & 0.76 ($\mp$ 0.18) & 0.9 ($\mp$ 0.09)\\
		BR1 $\longleftrightarrow$ BR3 & 0.57 ($\mp$ 0.21) & 0.92 ($\mp$ 0.08)\\
		BR1 $\longleftrightarrow$ BR4 & 0.65 ($\mp$ 0.22) & 0.93 ($\mp$ 0.07)\\
		BR1 $\longleftrightarrow$ BR5 & 0.8 ($\mp$ 0.15) & 0.99 ($\mp$ 0.03)\\
		BR2 $\longleftrightarrow$ BR3 & 0.77 ($\mp$ 0.16) & 0.95 ($\mp$ 0.07)\\
		BR2 $\longleftrightarrow$ BR4 & 0.8 ($\mp$ 0.16) & 0.94 ($\mp$ 0.07)\\
		BR2 $\longleftrightarrow$ BR5 & 0.87 ($\mp$ 0.12) & 0.98 ($\mp$ 0.05)\\
		BR3 $\longleftrightarrow$ BR4 & 0.63 ($\mp$ 0.2) & 0.91 ($\mp$ 0.09)\\
		BR3 $\longleftrightarrow$ BR5 & 0.75 ($\mp$ 0.17) & 0.97 ($\mp$ 0.05)\\
		BR4 $\longleftrightarrow$ BR5 & 0.76 ($\mp$ 0.17) & 0.96 ($\mp$ 0.07)\\

		\hline          	
	\end{tabular}
\end{table}

\begin{table}[!h]
	\caption{\textbf{Pearson's correlation coefficient between diversity indices (and the associated 95\% confidence interval in square brackets and significance level in parentheses) obtained with plant species and spectral cluster distributions within and between bioregions.} Significant correlations are marked with *** (p $<$ 0.001), while non-significant correlations are indicated as \enquote{ns}.}
	\label{TabS3}
	\centering
	\vspace{0.2cm}
	\begin{tabular}{lr}		
		\hline
		\\[-0.9em]
		\textbf{Bioregions and relationships} & \textbf{Pearson} \\ 		
		\hline	
		
		BR1 (H) & 0.109 [0.074,0.143] (***)\\
		BR2 (H) & 0.101 [0.062,0.139] (***)\\
		BR3 (H) & 0.038 [0.017,0.059] (***)\\
		BR4 (H) & -0.082 [-0.109,-0.054] (***)\\
		BR5 (H) & 0.17 [0.125,0.215] (***)\\
		All (H) & 0.072 [0.059,0.086] (***)\\
		BR1 ($\beta_{BC-bal}$) & 0.269 [0.268,0.27] (***)\\
		BR2 ($\beta_{BC-bal}$) & 0.302 [0.301,0.303] (***)\\
		BR3 ($\beta_{BC-bal}$) & 0.171 [0.171,0.171] (***)\\
		BR4 ($\beta_{BC-bal}$) & 0.256 [0.256,0.257] (***)\\
		BR5 ($\beta_{BC-bal}$) & 0.434 [0.432,0.435] (***)\\
		All ($\beta_{BC-bal}$) & 0.261 [0.261,0.261] (***)\\
		BR1 $\longleftrightarrow$ BR2 ($\beta_{BC-bal}$) & 0.139 [0.138,0.141] (***)\\
		BR1 $\longleftrightarrow$ BR3 ($\beta_{BC-bal}$) & 0.195 [0.194,0.195] (***)\\
		BR1 $\longleftrightarrow$ BR4 ($\beta_{BC-bal}$) & 0.015 [0.014,0.015] (***)\\
		BR1 $\longleftrightarrow$ BR5 ($\beta_{BC-bal}$) & 0.106 [0.105,0.108] (***)\\
		BR2 $\longleftrightarrow$ BR3 ($\beta_{BC-bal}$) & 0.184 [0.183,0.184] (***)\\
		BR2 $\longleftrightarrow$ BR4 ($\beta_{BC-bal}$) & 0.18 [0.18,0.181] (***)\\
		BR2 $\longleftrightarrow$ BR5 ($\beta_{BC-bal}$) & 0.239 [0.238,0.241] (***)\\
		BR3 $\longleftrightarrow$ BR4 ($\beta_{BC-bal}$) & 0.115 [0.115,0.116] (***)\\
		BR3 $\longleftrightarrow$ BR5 ($\beta_{BC-bal}$) & 0.27 [0.27,0.271] (***)\\
		BR4 $\longleftrightarrow$ BR5 ($\beta_{BC-bal}$) & 0.302 [0.301,0.303] (***)\\
		
		\hline          	
	\end{tabular}
\end{table}

\end{document}